\documentclass[journal, a4paper]{IEEEtran}

\usepackage{graphicx}   
\usepackage{url}  
\usepackage{amsmath}    

\begin{document}

% Define document title and author
	\title{Scaling Big Data Platform for Big Data Pipeline}
	\author{Rebecca Wild, Matthew Hubbell, Jeremy Kepner \\ MIT Lincoln Laboratory}
	\maketitle

% Write abstract here
\begin{abstract}
	Monitoring and Managing High Performance Computing (HPC) systems and environments generate an ever growing amount of data. Making sense of this data and generating a platform where the data can be visualized for system administrators and management to proactively identify system failures or understand the state of the system requires the platform to be as efficient and scalable as the underlying database tools used to store and analyze the data. In this paper we will show how we leverage Accumulo, d4m, and Unity to generate a 3D visualization platform to monitor and manage the Lincoln Laboratory Supercomputer systems and how we have had to retool our approach to scale with our systems.
\end{abstract}

 % Table~\ref{tab:name} 
 % Fig.~\ref{fig:name}.

\let\thefootnote\relax\footnotetext{This material is based upon work supported by the Assistant Secretary of Defense for Research and Engineering under Air Force Contract No. FA8721-05-C-0002 and/or FA8702-15-D-0001. Any opinions, findings, conclusions or recommendations expressed in this material are those of the author(s) and do not necessarily reflect the views of the Assistant Secretary of Defense for Research and Engineering.}
Leveraging the 3D Data Center Infrastructure Management (DCIM) tool built on the Unity game engine as published in 2015 \cite{1} has enabled the administrators of the TX-Green supercomputer at MIT Lincoln Laboratory Supercomputing Center, LLSC, to have an easily digestible single pane of the current state of the systems they manage. At the time of the original publication the TX-Green systems comprised of 3500 IT data points and 5000 environmental sensors outputs. The TX-Green systems were approximately 9,000 compute cores, 2PB of storage and a single core network. The integration of monitoring the core compute assets while having full situational awareness of the lights out data center located 100 miles from the offices was critical to providing a stable HPC platform for the research community. Enabling the administration team to proactively identify potential resource constraints, node failures, and environmental risks. 

The converged DCIM platform leverages the strategies and techniques commonly used in Big Data communities to store, query, analyze, and visualize voluminous amounts of data. It consist of Accumulo, MATLAB, and Dynamically Distributed Dimensional Data Model (D4M), and Unity \cite{2,3}. However, since original publication our systems have grown significantly. In 2016 we added a 1.32 Petaflop Intel Knights Landing system which debuted on the 2016 Top 500 list at 106 in the world \cite{4}. This addition brought over 40,000 additional compute cores to TX-Green, an additional core switch, and OmniPath network \cite{5}. As part of our regular refresh cycle we also included 75 GPU enabled systems to the TX-Green environment, thus creating a total of four unique computing architectures AMD, Intel, KNL, GPU requiring separate resource queues and creating more diverse landscape of available resources. The more than quadrupling of compute resources and service queues pushed our DCIM architecture to new levels.

Scaling our monitoring and management capabilities to match our new computational environment gave us good insight into the design choices and how they scaled under real world conditions. The underlying Accumulo database managed through the MIT SuperCloud portal technology has been the best performer, seamlessly scaling with the added entries and data collection fields \cite{6}. At the time of original publication the combined number of entries for both Node and Data Center databases was just over 15 billion. There are now 6.9 billion entries for the Nodes database and over 31 billion entries for the environmental building management system database. This would be extremely taxing on a standard mysql database. Accumulo, however, has performed exceptionally well under these conditions as it can withstand ingest rates of 100,000,000 entries per second \cite{7}.

 The scaling of our systems extended to all aspects of the HPC environment. With the additional computational resources being brought online and additional queues we expanded the number of default job slots individual users can consume on a single run from 2048 to 8192 and allowing some users to consume 16384 cores on special request. This results in the base number of jobs concurrently running on the system to be dramatically increased.

With additional queues setup for the four different architectures, new alerts were implemented to correlate to the heterogeneous environment of available memory, local disk, and CPU load thresholds. As a result the growth in data collections for each node grew substantially. The 40,000+ cores added to the system were each reporting the jobs running, the potential alerts and thresholds met. The one area of our processing pipeline that was most affected by this explosion in data was the Unity visualization platform. 

\begin{figure}[!hbt]
		\begin{center}
		\includegraphics[width=\columnwidth]{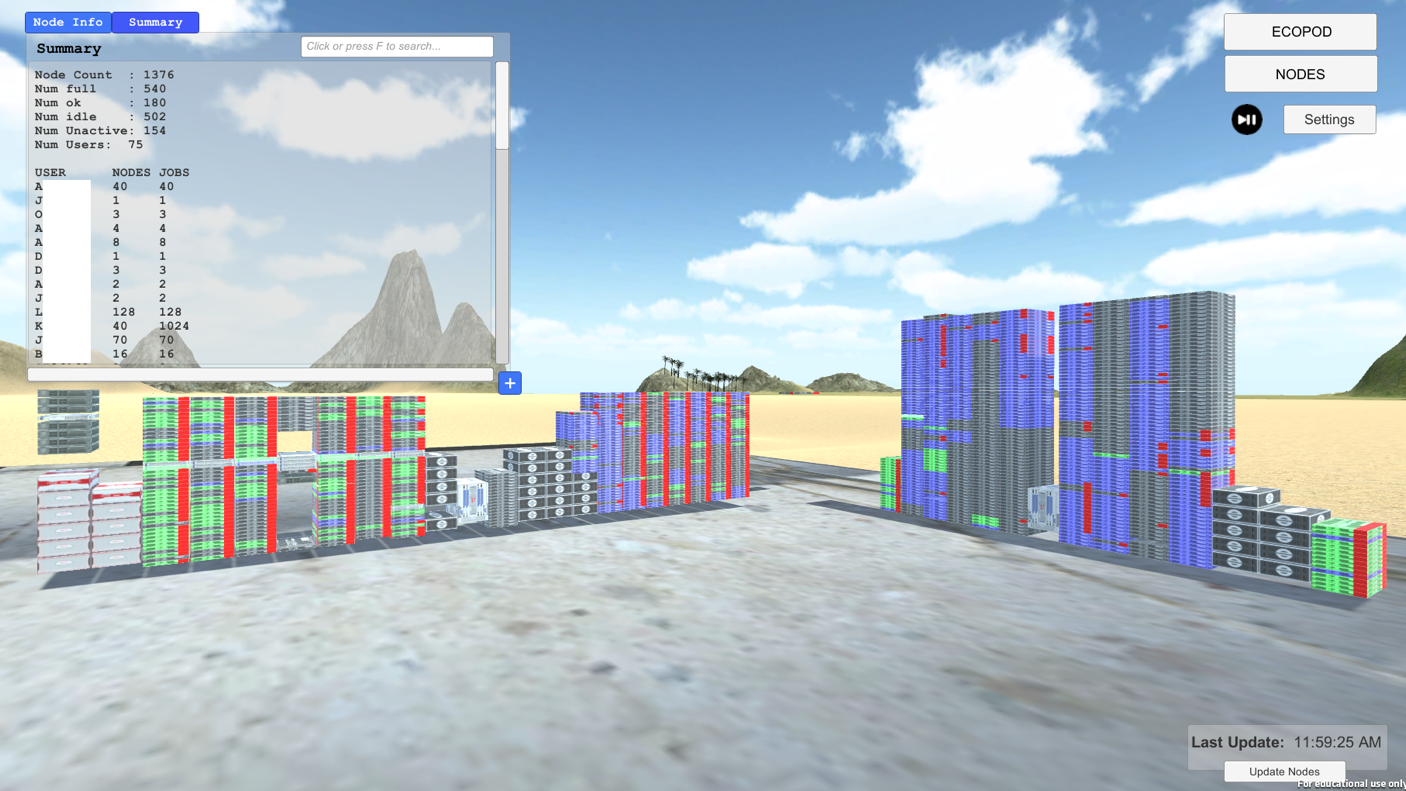}
		\caption{User perspective of the new visualization environment.}
		\label{fig:nodeView}
		\end{center}
	\end{figure}

The rendering of the additional nodes and game objects in the 3D space was not impacted, however, when applying the data to the nodes and subsequent updates the visualization environment performance dropped off significantly. On every update cycle the load time would stall 10+ seconds to parse the .csv files and apply the updated data to the EcoPOD and nodes. We applied a number of strategies to try address the constant lagging. First, we decided to stagger the updates of the EcoPOD and the nodes so the call was less expensive and hopefully less noticeable to the user. This only led to multiple slightly shorter lags in the game environment and a more stuttered playing experience. Secondly, we tried to chunk the updates so only a subset of the nodes would update at a time. Unfortunately, this also did not resolve the issue as the environment was constantly updating and the lags, while smaller, were more frequent. We finally succumbed to the fact that we had been carrying on too much legacy code and Unity has made many updates since the 3.0 version we had originally built the platform on. We decided to do a ground up rewrite of the environment with more attention to scale and performance Fig.~\ref{fig:nodeView}. 

    % Cores v Jobs Graph

The original code was primarily written in UnityScript, Unity's JavaScript-like language available as an alternative to C\#. UnityScript however has no tangible advantage, little documentation, and is in the deprecation process \cite{unityScript}. For these reasons we opted to rewrite the environment using C\#. 

A major fault of the original environment was that it was too modular. Every node was a game object with 4 scripts attached each inherent from Unity's MonoBehavior class meaning the script has an update function called by the game engine every frame. This is a clear source of overhead. Each frame the node would check to see if their was new data ready to be processed. When new data was available it retrieved this data and updated it's attributes and appearance. Under this model all nodes updated concurrently causing the game engine to spend too much time switching contexts resulting in long update times.

\begin{figure}[!hbt]
		\begin{center}
		\includegraphics[width=\columnwidth]{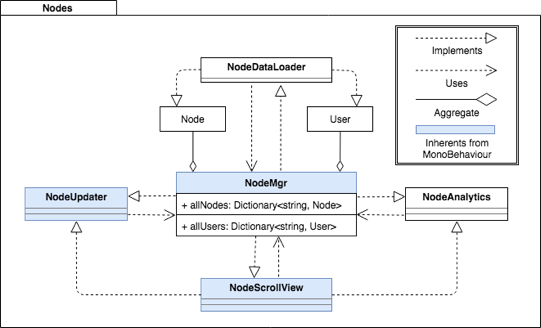}
		\caption{Node Data Management Design Model}
		\label{fig:nodeUML}
		\end{center}
	\end{figure}

In the rewrite we redesigned the update procedure so that everything would happen sequentially. Instead of having each node update itself concurrently we created a script that to manage all of the node updates reducing the number of scripts attach to each node from four to zero, greatly reducing the number of update calls per frame. When new data is ready to be received the managing script, \verb|NodeMgr|, uses utility classes shown in Fig.~\ref{fig:nodeUML} to load the new node data into a Dictionary data structure, update the nodes appearance, update the information on the display panel, and update the analytics of the system state. By preforming a single update task on each node sequentially we enforce spacial and temporal locality and are able to better achieve the appearance of synchronization than if the nodes were to update concurrently. A very similar design model is used to update the EcoPOD.

	% Node UML Diagram Figure
	
The use of Dictionaries to store the data in key value format and act as the aggregator to then query from within the game. It allows us to pull the necessary request from either the end user or the in-game analytics on demand. It also has a constant look up time, ensuring us speed with scalability.

To retrieve data from the nodes or EcoPOD components we use a raycast from the mouse pointer to identify the game object, get its name, look it up in the Dictionary, and retrieve the corresponding data.  By using this approach, we eliminated the need for each node to preform physics calculations for collision detection with the mouse pointer or surrounding game objects. This reduced the time spent calculating physics and processing the huge push of data from the input files directly to the assets in the game environment. The end result was a reduction of over 5,000 method calls per frame during an update to just three. This led to a dramatic improvement in performance:

	% Performance Improvements Table
	\begin{table}[!hbt]
		\begin{center}
		\caption{Performance Improvements}
		\label{tab:performanceImprov}
		\begin{tabular}{|c|c|c|}
			\hline
			 & Original M\&M & Updated M\&M \\ \hline
			EcoPOD Update & $11.26$s & $0.55$s \\ \hline
			Node Startup & $27.19$s & $0.59$s \\ \hline
			Node Update & $10.76$s & $0.65$s \\ \hline
		\end{tabular}
		\end{center}
	\end{table}

The performance was maintained even when we doubled the in-game assets reflecting future expansions of an additional data center and next generation supercomputing systems.

	% Scaling Test Results Table
	\begin{table}[!hbt]
		\begin{center}
		\caption{Scaling Test Results}
		\label{tab:scalingResults}
		\begin{tabular}{|c|c|c|}
			\hline
			 & Updated M\&M & Updated M\&M x2 \\ \hline
			EcoPOD Update & $0.55$s & $0.58$s \\ \hline
			Node Startup & $0.59$s & $0.68$s \\ \hline
			Node Update & $0.65$s & $0.90$s \\ \hline
		\end{tabular}
		\end{center}
	\end{table}

By reorienting the data flow to be stored in memory and using traditional efficient data structures to pull the data to the game objects rather than pushed in a single update opened up the platform to many more analytic possibilities. 

By redesigning the data flow we were able to create an efficient and scalable visualization platform granting us many more analytic possibilities. Unity recently added support for AI and Machine learning agents \cite{8}. We can foresee the ability for the game to identify conditions that could lead to component failures and alerting the administration. This area of research is beginning to pick up traction with the recently reported collaboration of nlyte, a leading DCIM provider, and IBM Watson to being identifying areas for operational efficiency \cite {9}.

% Bibliography

% Your document ends here!
\end{document}